# Improved Sensing and Positioning via 5G and mmWave radar for Airport Surveillance

Bo Tan*, Elena Simona Lohan*, Bo Sun*, Wenbo Wang*, Taylan Yesilyurt*, Christophe Morlaas†,
Carlos David Morales Pena†, Kanaan Abdo‡, Fathia Ben Slama‡, Alexandre Simonin‡, Mohamed Ellejmi§
*Faculty of Information Technology and Communication Sciences, Tampere University, Finland;
†ENAC/TELECOM, University of Toulouse, ENAC, Toulouse, France;
‡ALTYS Technologies, Toulouse, France; §EUROCONTROL, Bretigny Sur Orge, France
{bo.tan, elena-simona.lohan, bo.sun, wenbo.wang, taylan.yesilyurt}@tuni.fi,
{christophe.morlaas, carlosdavid.moralespena}@enac.fr
{kanaan.abdo, fathia.benslama, alexandre.simonin}@altys-tech.net, mohamed.ellejmi@eurocontrol.int

*Abstract*—This paper explores an integrated approach for improved sensing and positioning with applications in air traffic management (ATM) and in the Advanced Surface Movement Guidance & Control System (A-SMGCS). The integrated approach includes the synergy of 3D Vector Antenna with the novel time-of-arrival and angle-of-arrival estimate methods for accurate positioning, combining the sensing on the sub-6GHz and mmWave spectrum for the enhanced non-cooperative surveillance. For the positioning scope, both uplink and downlink 5G reference signals are investigated, and their performance is evaluated. For the non-cooperative sensing scope, a novel 5Gsignal-based imaging function is proposed and verified with realistic airport radio-propagation modelling and the AI-based targets tracking-and-motion recognition are investigated. The 5G-based imaging and mmWave radar-based detection can be potentially fused to enhance surveillance in the airport. The work is being done within the European-funded project NewSense and it delves into the 5G, Vector Antennas, and mmWave capabilities for future ATM solutions.

*Keywords*—5G, Angle of Arrival, Air Traffic Management, Communications, Millimetre Wave, Positioning, Sensing, Surveillance, Time of Arrival, Vector Antenna, Synthetic-aperture Radar.

## I. Introduction And Motivation

Innovative technologies - not traditionally specific to Air Traffic Management (ATM) domain - are of increasing interest in order to support, in a sustainable and efficient way, the various Communication, Navigation, and Surveillance (CNS) requirements in airport areas. Tracking solutions based on timing and angle measurements from 5G radio frequency (RF) signals, mmWave radar, and 5G RF-based radar (i.e. using 5G signals for radar/sensing purposes) have not been addressed so far in the ATM context to the best of the Authors' knowledge. Yet, there is a significant technological convergence opportunity of future ATM with communication and navigation solutions based on terrestrial 5G networks. The 5G current spectrum lies around 3.5 GHz and future 5G and 6G systems will also operate in mmWave bands (near 26 GHz and above 60GHz), falling outside the current aviation spectrum, and therefore having the promise of low interference with existing ATM signals and of enabling complementary solutions to those currently existing in ATM. The anticipation of large-scale deployment of 5G networks will allow detection of all sorts of objects in the airport area including unmanned aerial vehicles (UAV) and will also enable accurate positioning of terrestrial and UAV targets, by taking advantage of the richness of reference signals available nowadays in 5G systems [1]–[4]. Radar solutions based on mmWave, widely used for automotive industry, having compact size including antennae, with very competitive cost compared to traditional radars, and providing reliable performance, could also be used to track and classify [5] non-cooperative targets at airport surface.

Most of the world's large airports nowadays are already deploying Advanced Surface Movement Guidance and Control System (A-SMGCS) surveillance service, implementing multilateration (MLAT) technologies, alongside with traditional surface movement radar (SMR). However, the situation for small and medium-sized airports is quite different: on one hand, the air traffic growth puts the secondary airfields under pressure for greater capacity and increased safety, but on the other hand, the current surface surveillance technologies remain out of their reach because of prohibitive infrastructure costs [6]. Being able to re-use existing infrastructure, such as the 5G networks - expected to be widely deployed worldwide and the mmwave radar used for automotive and industrial applications, would help towards providing technically and financially viable solutions for surveillance service for small and secondary airports. Thus we propose a collaborative approach for the small and medium-sized airports surveillance as illustrated in Fig1. This concept relies on innovative and low-cost surveillance solutions developed outside the ATM Surveillance Industry, for example 5G-signals-based solutions. We rely on the assumption that large-scale 5G deployments will be available worldwide in the next decades, with independent private 5G networks available at airports and 5G equipment on-board of aircraft.

By taking advantage of the pre-defined uplink (UL) sounding reference signal (SRS), downlink (DL) channel status information reference signal (CSI-RS), and DL positioning reference signal (PRS), the joint angle-delay estimation methods [7], [8] can be applied with the array receiver to localize the 5G user equipments (UEs) in UL and 5G base stations (called gNBs) in DL. Besides the positioning via the reception of the direct transmitted 5G signal, we also exploit the echos of the reference signals for the imaging sensing purpose by mono-static Synthetic-aperture radar (SAR), in UL scenario and bi-static SAR in DL scenario.

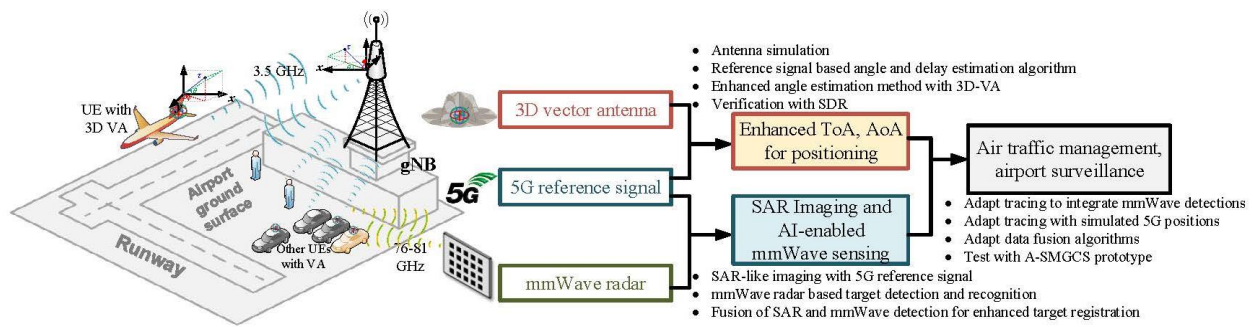

Figure 1. Illustration of the integrated-concept architecture for airport surveillance

Under the 5G framework, 3D vector antenna (3D-VA) and mmWave radar are also adopted to enhance the positioning and sensing functions needed in the ATM and airport surveillance. The combination of the 3D-VA and 5G signal will lead to the improved angle resolution because of the polarization differentiation capability brought by orthogonal electric-magnetic reception from the three axes of a 3D-VA. The fusion of the targets in the 5G-signal-based imaging and mmWave-radar detection enhances the target detection and recognition in the surveillance scenarios. The integration of the 5G reference signal, 3D-VA, and mmWave radar in this work is tackled from the following three points of view:

- **Single-cell UE Positioning:** Joint delay and angle localization approaches are proposed for single 5G gNB scenario. Estimation methods utilise SRS or PRS transmitted between the UE and gNB, and receivers are equipped with VA and URA;

- **VA Array enhanced angle estimation performance:** Leveraging the antenna array to compose 3D-VA array to estimate the UE's angle from SRS transmitted from flight to the gNB;

- **Joint positioning and identification:** Taking the advantage of the unique reference sequence allocated to individual 5G UE in the network, the 5G signal based solution provides the location naturally together with the identification information without support from auxiliary system. This fact potentially leads to an economic ATM/ATC solution for the futue secondary airports.

- **Complementary Sensing:** The 5G-signal-based imaging provides an opportunity to identify the non-cooperative targets from the SAR images. The mmWave radar provides more accurate position estimation and detailed motion status than the sub-6GHz 5G-signal-based imaging/sensing. With a proper target-association strategy, the detection and motion recognition accuracy can be further improved by using hybrid radio signal sources.

The enhanced positioning and complementary sensing functions show promising potential to provide the economic solutions for the future ATM and airport surveillance. EUROCAE has defined performance requirements for Mode S Multilateration (MLAT) Systems and SMR sensor systems for use in A-SMGCS in ED-117 [9] and ED-116 [10] respectively. As the sensors developed in the scope of current study (5G reference signal combined with 3D-VA and mmWave radar) come as alternatives to costly technologies stated before, they will be assessed using the performance requirements as defined in these standards when applicable. A preliminary requirement analysis has been already defined in Deliverable 1.1 of the NewSense [11], for example, Reported Position Accuracy is $\leq$ 12 m (95%) for 5G surveillance sensors in the Manoeuvering Area and it is $\leq$ 7.5 m (95%) for mmWave radar. According the 5G standard, the positioning accuracy will need to achieve 3 to 10 m for 80% for outdoor scenarios in Release 16 [12], further narrow down to 0.3 m and with 10 ms positioning latency in Release 17 [13].

In the rest of this paper, we will introduce the working principle of the 3D-VA in Section II; in Section III the 5G reference signals, 3D-VA involved angle-delay estimation algorithms, and the imaging algorithms are elaborated in detail; the AI-based mmWave radar sensing and target recognition are described in Section IV. In Section V, we provide the preliminary results of the positioning, imaging, and target recognition. At last, the conclusion and future work are given in the Section VI.

## II. DOA VIA 3D VECTOR ANTENNA

### A. Vector antenna presentation

A theoretical vector antenna (VA) is composed of six orthogonal and collocated antennas that combine three electric and three magnetic elementary dipoles, as illustrated in Fig. 2 (a). These six dipoles allow the derivation of the Direction of Arrival (DoA) from the measurement of the six components of the $k^{th}$ incident Electromagnetic (EM) field in the Cartesian coordinate system, whatever the polarization state. Its DoA is defined by the azimuth $\phi_k$ and elevation $\theta_k$ angles. The main advantages of such a VA antenna are wider angular coverage (theoretically the full 3D space) and compact size (one wavelength).

### B. Proposed vector antenna

The proposed VA is made of two vertical parts and one horizontal part as illustrated in Fig. 2 (b). Each vertical part is orthogonal to each other and consists of a dual-port semicircular arrays of four Vivaldi antennas. The horizontal part consists of a circular array of eight Vivaldi antennas with four ports. Therefore, this VA has a total of eight ports: four ports (ports numbers from 1 to 4) are associated with the vertical part; and 4 ports (ports numbers from 5 to 8) are associated with horizontal part. The vertical part enables the measurement of three components of the vertically-polarized incoming EM field (namely $E_Z$, $H_X$, and $H_y$) while the horizontal part permits the measurement of the other three components (namely $H_Z$, $E_X$, and $E_Y$) of the horizontally-polarized incoming EM field.

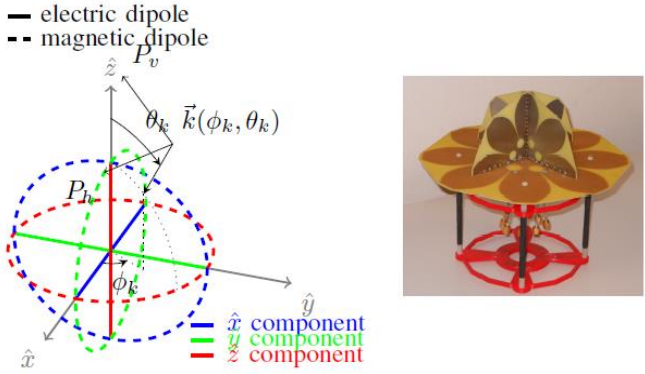

Figure 2. (a) Topology of a theoretical VA; (b) Photograph of the 8-port vector

## C. Observation model

For $k$ EM-fields incident upon the VA, the voltage $Y \in \mathbb{C}^{M,1}$ received at time t on each M ports of the VA can be written as follows:

$$\mathbf{Y}(t) = \mathbf{D}(\phi,\theta,\gamma,\eta) \cdot \mathbf{s}(t) + \mathbf{N}(t) \quad (1)$$

where $\cdot$ is the dot-product function, $D \in \mathbb{C}^{M,k}$ is the angle of arrival (AoA) steering vector corresponding to the VA response including the polarization vector $\boldsymbol{p}$ and the spatial transfer function $\boldsymbol{F}_s$.

$$\mathbf{D}(\phi,\theta,\eta,\gamma) = \mathbf{F}_s(\phi,\theta)\mathbf{p}(\eta,\gamma) \quad (2)$$

The polarization vector $\boldsymbol{p}$ is defined as the combination of vertical ($P_v$) and horizontal ($P_h$) polarization states as follows

$$\mathbf{p}_k = \sin(\gamma_k) e^{j\eta_k} P_v + \cos(\gamma_k) P_h \quad (3)$$

where $\gamma_k \in [0°; 90°]$ and $\eta_k \in [-90°; 90°]$ refer to the auxiliary polarization angle and the polarization phase difference, respectively.

Moreover, $\boldsymbol{s} \in \mathbb{C}^{K,1}$ designates the signal vector associated with the $k$ incoming EM-fields at time t and $\boldsymbol{N} \in \mathbb{C}^{M,1}$ denotes the additive white Gaussian noise. It is assumed here that this noise is spatially invariant with zero-mean and covariance matrix $\boldsymbol{R_n} = \sigma_n^2 \mathbf{I} \in \mathbb{C}^{M,M}$.

Then, for an ideal VA, $\boldsymbol{F}_s$ can be modeled by the matrix describing ideal elementary dipoles without coupling.

$$\begin{bmatrix} \cos\phi\cos\theta & -\sin\phi \\ \sin\phi\cos\theta & \cos\phi \\ -\sin\theta & 0 \\ -\sin\phi & -\cos\phi\cos\theta \\ \cos\phi & -\sin\phi\cos\theta \\ 0 & -\sin\phi \end{bmatrix}$$

For the proposed VA, the spatial transfer function is deduced from EM simulation or measurement when the VA is illuminated with each incoming EM wave defined by its direction ($\phi_k$, $\theta_k$), its polarisation state ($\eta$, $\gamma$) at one frequency, where $\phi_k \in [0°; 360°]$ and $\theta_k \in [0°; 180°]$.

## III. SENSING & POSITIONNING IN SUB-6GHZ B

### A. Reference Signals in 5G

In 5G, a variety of reference signals are used for many different purposes such as channel sounding, synchronization, demodulation, etc. By using these signals, different UL and DL positioning methods can be implemented in the location server. In this work, three different reference signals frequently used for positioning are investigated, namely: SRS for the UL transmissions and CSI-RS and PRS for the DL transmissions.

*1) SRS:* This signal is the UL reference signal, periodically transmitted from UE to gNB to enable UL channel sounding. Zadoff-Chu sequences are applied to SRS resource elements, due to their specific properties. SRS can be extended to support better positioning. The duration in time can be extended up to 12 OFDM symbols. In addition, comb size ($K_{TC}$) of an SRS can be increased up to 8. Therefore, more UEs can be multiplexed [15]. Fig. 3 illustrates two possible SRS configurations.

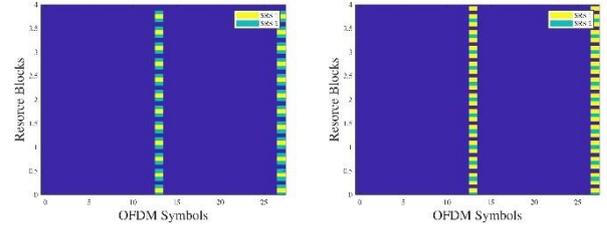

Figure 3. Multiple SRSs multiplexed in the same time slot with different comb size: a) $K_{TC1}$=4, $K_{TC2}$=2; b) $K_{TC1}$=2, $K_{TC2}$=4

*2) CSI-RS:* This DL-only reference signal is a signal generated by a length-31 pseudo-random sequence defined in 3GPP TS 38.211 (section 5.2.1). For each configured CSI-RS, the UE shall assume the reference signal sequence being mapped to resources elements $(k,l)_{p,m}$ so that the UE can assume a configured CSI-RS signal will not collide with any signal configured for the device, such as demodulation reference signals associated with DL transmissions [15].

*3) PRS:* is another DL reference signal specified in 3GPP Release 16 to support DL positioning methods. Similarly with CSI-RS, this DL-only reference signal is a signal generated by a pseudo-random sequence. Different comb sizes (2,4,6,12) can be configured to a PRS. Therefore, multiple base stations may use the same time slots, but different sets of subcarriers can be multiplexed without interference. In addition, PRS signals can be muted according to a muting pattern known by UE to avoid interference more effectively.

SRS is used for UL positioning, while PRS and CSI-RS are used for DL positioning in 5G.

### B. Angle and Delay Estimation Methods

Let us assume that the received signal $Y^m(t)$ from the mth UE experiences the propagation delay t with direction parameters φ and θ; we assume also that the source is linear polarised, and transmitter is vertically placed, so $\gamma$ and $\zeta$ equal to 90˚ and 0˚. Now, we have delay steering vector $\mathbf{g}(t)$ in (4) and AoA steering vector $\mathbf{D}(\varphi,\theta)$.

$$\mathbf{g}(t) = [1, e^{-2j\pi f_1 t}, e^{-2j\pi f_2 t}, ..., e^{-2j\pi f_{n-1} t}] \quad (4)$$

After modelling the steering vectors, they are used in Subspace-based and Expectation–maximization (EM) methods for positioning parameters estimation.

*1) Subspace Method:* It is based on the subspace algorithm Multiple Signal Classification (MUSIC) [16]. This approach firstly calculates the covariance matrix of the received signal $Y^m(t)$ from the $m$th UE.

$$R_{YY} = \mathbb{E}[Y^m(t)Y^m(t)^*] \quad (5)$$

where $\mathbb{E}[\cdot]$ denotes expectation, $[\cdot]^*$ means the conjugate transpose. Then, it applies the eigenvalue decomposition to get $\lambda = [\lambda_1, \lambda_2, \ldots, \lambda_{6N}]$ (with ascending order) and eigenvector $E_n = [e_1, e_2, \ldots, e_{6N}]$. By eliminating the $K$ strongest values from $E_n$, we can get the noise subspace $E_n = [e_1, e_2, \ldots, e_{6N-K+1}]$. As we only estimate Line of Sight (LoS) path, $K = 1$ in this work.

Then, the parameters corresponding to the peak value of $P(\varphi, \theta, t)$ in (6), with brute force search, indicate the received signal delay and AoA, respectively.

$$P(\phi, \theta, t) = \frac{A^*(\phi, \theta, t)A(\phi, \theta, t)}{A^*(\phi, \theta, t)E_n E_n^* A(\phi, \theta, \gamma, t)} \quad (6)$$
$$A(\phi, \theta, t) = D(\phi, \theta) \otimes g(t)$$

where $\otimes$ is Kronecker product.

*2) EM Method:* This method follows the space-alternating generalized expectation-maximization (SAGE) design in [8]. It repeats Expectation step (E-step) and Maximization step (Mstep) to minimise the difference between the expected signal and the observed signal. E-step firstly assumes some delay and AoA parameter values to generate $\hat{Y}^m(t)$ as (7) showed. We the noise component $\hat{N}^m(t)$ equals to $Y^m(t) - \hat{Y}^m(t)$. Then, M-step exhaustively searches $L(\varphi, \theta, t)$ in $t$, $\varphi$ and $\theta$ three dimensions as (8) expressed. Parameters which produce the peak value of $L(\varphi, \theta, t)$ will be used in the next round E-step to generate a new $\hat{Y}^m(t)$ to compare with the observed signal $Y^m(t)$. The iteration process will stop when the difference between $\hat{Y}^m(t)$ and $Y^m(t)$ is smaller than a predefined convergence value $\xi$.

$$\hat{Y}(t) = D(\phi, \theta)s(t) + \hat{N}(t) \quad (7)$$
$$\hat{N}(t) = Y(t) - \hat{Y}(t)$$

$$L(\phi, \theta, t) = D^*(\phi, \theta)\hat{Y}^m(t)g^*(t) \quad (8)$$

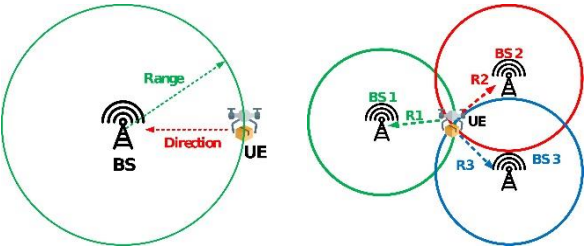

Figure 4. (a) Single Base Station Positioning Scenario; (b) Multi-Base Station Positioning Scenario

## C. Single and Multiple-Cell Approaches

In single base station case, joint ToA/AoA positioning problem becomes an angle-of-arrival and delay-estimation problem. Delay is the time difference between transmission time and reception times. From the product of the delay with propagation speed, the range (r) is calculated. As seen in Fig. 4 (a), the position of UE is shown as the intersection of the red line indicating the direction of the received signal and green circle indicating the range. After estimating the angle and range, where the location of UE ($X_u, Y_u, Z_u$) can be calculated by sum the relative coordinates with BS's coordinate as (9) expressed.

$$X_u = X_{BS} + r\cos(\theta)\sin(\phi)$$
$$Y_u = Y_{BS} + r\sin(\theta)\sin(\phi) \quad (9)$$
$$Z_u = Z_{BS} + r\cos(\phi)$$

Using multiple base stations (BS) for ToA and AoA positioning may overcome some of the challenges encountered in single BS positioning. In Fig. 4 (b), two additional base stations build LoS communication with UE to complement the single base-station situation from Fig. 4(a). The simplest way to reduce the positioning error is to average coordinates estimated from multiple BSs (or gNBs). Equation (10) shows the averaging result of $X_u^{BS1}, X_u^{BS2}$ and $X_u^{BS3}$ from three BSs.

$$X_u^{averaging} = 1/3 * \sum_{i=1}^{3} X_u^{BS(i)}$$
$$X_u^{BS(i)} = X_u^{GT} + X_{error}^{BS(i)} \quad (10)$$

where $X_u^{GT}$ is the ground truth of UE's coordinate in x-axis. Measurement errors $X_{error}^{BS(i)}$ from multiple BSs are averaged to reduce the high measurement error caused by a single BS.

## D. 5G Signal Based Imaging

Besides using the reference signals for positioning purpose to estimate angle and delays, one can also exploit the reflection of the reference signal for sensing functionality. In this paper, we elaborate an image-function based on UL SRS signal to mimic the synthetic-aperture radar (SAR) mechanism. By using the UL SRS from the UE (e.g., airplane) as the illuminator, a monostatic SAR system can be formed by collecting and processing the reflected signals from ground objects such as buildings and vehicles. Range-Doppler information carried by the reflections from objects on the ground is directly linked with the geometry of those targets, and the usage of 5G reference signal enables us to achieve this function under 5G networks. In Fig. 5, we assume a moving aircraft is capable of collecting echoes of SRS from the ground. The down range and cross range are linked with propagation delay and Doppler frequency shift. For the imaging purpose, we assume the reflected signal is $Y_{RS} \in \mathbb{C}^{n,K}$ with n subcarriers and K OFDM symbols. The delay and Doppler information of $L$ reflections can be written as follows:

$$C^i(v_l) = e^{-2j\pi i T_{cpsym} 2v_l/\lambda} \quad (11)$$

$$\mathbf{D}(v_l) = [1, e^{-2j\pi T_{sym} 2v_l/(n\lambda)}, ..., e^{-2j\pi(n-1)T 2v_l/(n\lambda)}] \quad (12)$$

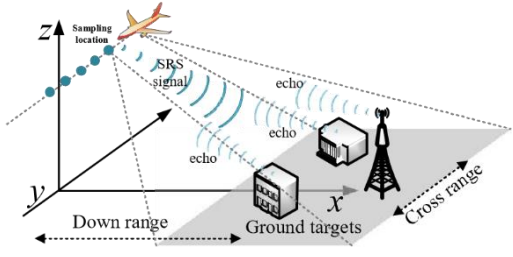

Figure 5. An example of the airport surveillance scenario with 5G uplink SRS signals.

For every path $l$, $\mathbf{C}^i(\mathbf{v}_l)$ and $\mathbf{D}(\mathbf{v}_l)$ are Doppler shift notation on $i$ th symbol and all subcarriers. $v_l$ is the Doppler speed and wavelength is $\lambda$. Duration of OFDM symbol and cyclic prefix (CP) OFDM symbol are $T_{sym}$ and $T_{cpsym}$, respectively. The received $i$th reflected symbols is:

$$\mathbf{Y}_{RS}^i = \sum_{l=1}^{L} C^i(v_l)\mathbf{D}(v_l) \odot \mathbf{g}_i(t_l) \odot \mathbf{X}_{RS}^i + \mathbf{N}_{RS}^i \quad (13)$$

Where $\mathbf{X}_{RS}^i$ is the frequency domain transmitted sequence. For extracting Doppler information, M OFDM symbols are collected to construct the data matrix $\mathbf{Y}_{RS}$.

$$\mathbf{Y}_{RS} = [\mathbf{Y}_{RS}^{1\ T}, \mathbf{Y}_{RS}^{2\ T}, ..., \mathbf{Y}_{RS}^{MT}] \quad (14)$$

We multiply the $\mathbf{Y}_{RS}^i$ with $\mathbf{X}_{RS}^{*i}$, conjugation of transmitted signal, to get the matched filter output $\mathbf{H}_{RS}^i$.

$$\mathbf{H}_{RS}^i = \mathbf{X}_{RS}^{*i} \times \mathbf{Y}_{RS}^i \quad (15)$$

$$\begin{aligned}
\mathbf{f}(\tau, v) &= (\mathbf{g}^*(\tau) \odot \mathbf{D}^*(v))\mathbf{H}_{RS}\mathbf{C}^*(v) \\
\mathbf{H}_{RS}^i &= [\mathbf{H}_{RS}^{i\ T}, \mathbf{H}_{RS}^{i\ T}, ..., \mathbf{H}_{RS}^{i\ T}] \\
\mathbf{C}(v) &= [C^1(v), C^2(v), ..., C^M(v)]^T
\end{aligned} \quad (16)$$

After searching the peak value along speed v and delay τ two dimensions in (16), we can generate the range Doppler plot of the illuminated area. The visualized process can be seen in Fig. 7.

## IV. SENSING ON THE MMWAVE BAND

### A. mmWave-Radar Overview

The mmWave radar relies on highly directional electromagnetic Frequency Modulated Continuous Waves (FMCW), operating typically at 24GHz and 77−81GHz, and mitigating environmental factors, as they are not affected by heat or light [17]. They can be also tuned for short, long, and widedetection ranges to meet airport surface surveillance needs. Fig. 6 illustrates the functional architecture of the FMCW radar with multiples receiving antennae (Rx1, Rx2 ...RxN). The FMCW radar emits a high-frequency signal called a chirp, with a frequency that increases linearly with a certain slope s during the measurement phase. Once the chirp is reflected by an object on its path, the reflected chirp is received by the receive antennae (e.g. Rx1). Then, for each Rx, the Rx and Tx signals are mixed, to generate the Intermediate frequency (IF) signal. This obtained signal is then filtered with a low-pass filter to eliminate the high frequency components, to be next digitized using the Analog to Digital Converter (ADC). ADC samples are then processed to calculate target's range, velocity and angle through multiple calculations mainly based on Fast Fourier Transform (FFT). Signals patterns can also be used to recognize object type using AI classifier [5] [18] [19].

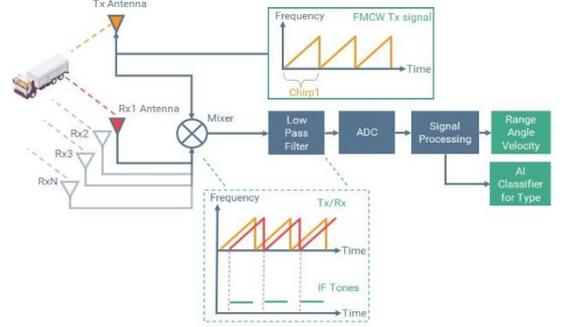

Figure 6. mmWave radar architecture with multiple Rx Antenna

### B. Signal Processing Calculations

*1) Range FFT Calculation:* A first, the FFT calculation applied directly on ADC data (resulting from each chirp) results in a frequency spectrum (Fig. 15) that has separate peaks. Each peak, at frequency $f_{peak}$, denotes the presence of an object at a specific distance.

$$distance = \mathbf{f}_{peak} \frac{\mathbf{c}}{2 \times \mathbf{s}} \quad (17)$$

This calculation is named, 1D FFT processing (also named range-FFT).s in (17) is frequency shift per unit of time. This calculation is applied on every chirp.

Multiple objects equidistant from the radar, but with differing velocities relative to the radar, will have single peak corresponding to the range. These equi-range objects which have different velocities relative to the radar can be separated using a Doppler-FFT named 2D FFT.

*2) Doppler/Velocity FFT calculation:* The range-FFT corresponding to each chirp will have peaks in the same location, but with a different phase. The measured phase difference corresponds to the moving velocity. The 2D (velocity) FFT processing that takes as input the 1D FFT and performs FFT to give a (range, velocity) matrix is illustrated in Fig. 7.

*3) Angle FFT calculation:* The 3D FFT processing for direction of arrival (azimuth) estimation is used to map the X-Y location of the object. The range-FFT and Doppler-FFT do not work when two objects with equidistant and same velocity relative to the radar appear. Then the AoA needs to be estimated. The AoA estimation is based on the phase change in the peak of the range-FFT or Doppler-FFT, because of differential distance from the object to each of the antennas, which requires at least two Rx antennae. Similarly, an FFT on the sequence of phasors corresponding to the 2D-FFT (range-FFT and Doppler-FFT) peaks resolves the angle estimation problem. This is called angle-FFT.

### C. Clustering and tracking

Detection's (range, angle equivalent to X,Y coordinates) resulting from FFT calculations are then clustered to identify,

track, and map detected targets. Clustering consists of finding and removing outliers from the data that will be used for tracking and objects classification which will improve the performance of objects detection's. A well-known data clustering algorithm that is typically used in Machine Learning (ML) is the Density-Based Spatial Clustering of Applications with Noise (DVBSCAN) [20]. DBSCAN clusters the points that are close to each other in a specific group based on two parameters: *i)*. The distance from an original point to surrounding points (named eps); and *ii)*. The minimum number of points in one circle which center is the original point and radius eps. The points that satisfy these two criteria are clustered in the same core. The points that do not satisfy these two parameters are classified as border and outlier (noise) points.

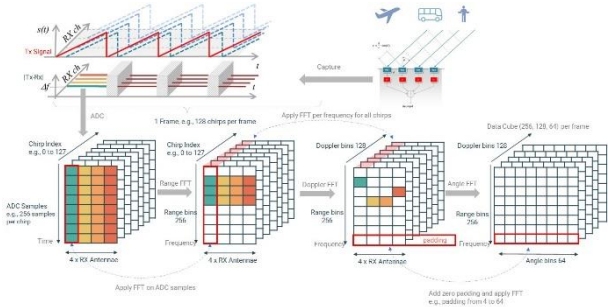

Figure 7. mmWave signal processing flow and calculations

### D. Target's classification using AI

The use of AI combined with cameras and/or LIDAR for positioning and classification has been already studied and developed for use in A-SMGCS. However, this solution is affected by environmental conditions, such as rain, dust, smoke, fog, frost, darkness, and direct sunlight. mmWave radar integrated with AI enables to overcome these limitations as the positioning (using signal processing functions listed above) and targets classification using Machine Learning will be based on the radar signal. Objects' classification identifies the target type, for airport context, three classes are targeted: aircraft, vehicle, and persons. Multiple studies proposed the use of ML in radar signal detection and classification [5] [19] [21].

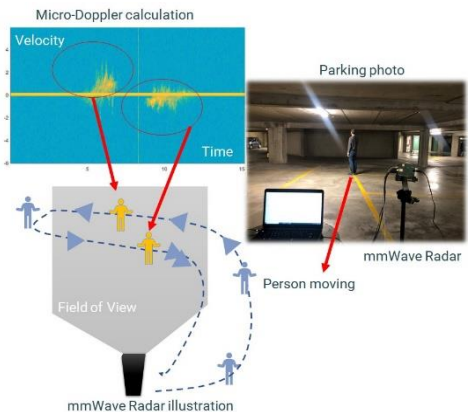

Figure 8. MD plot showing velocity signature of a person in time

mmWave radar calculations are used as input to the ML model including radar point clouds, 2D-FFT, 3D-FFT, and Micro-Doppler (MD) signatures. A MD Signature [22] of a target is a time varying frequency modulation imparted on the radar echo signals by moving components of the target. The following Fig. 8 illustrates a measurement performed in a parking for a person walking forth and back in front of the mmWave radar. The micro Doppler generates a certain signature that will be used to train a model on the target type.

## V. PRELIMINARY RESULTS AND DISCUSSION

### A. 3D VA Results

The antenna presented in [14] has been optimized for 5G signals. This antenna is enable to cover the all sub-6GHz 5G band from 1.9GHz to 6GHz. The larger dimension is the horizontal part with 24cm. The impedance bandwidth is presented in Fig. 9 (a), and it is from 1.3GHz to 9GHz with reflection coefficient $S_{ii} < -10dB$ and specifically $S_{ii} < -20dB$ in the optimized bandwidth from 1.9 to 3.5GHz. We remark only one port of the vertical and horizontal antenna part is represented. The other ports have equivalent performances due to antenna symmetry. In Fig. 9 (b), the isolation coefficient between ports is plotted. A mutual coupling reduction of $-20dB$ is obtained for each port allowing a good orthogonality properties between ports. In term of radiation pattern an example is given in Fig. 10 and 11 for an excitation state corresponding to the electric and magnetic dipole respectively, along $z$ axis. Good polarization discrimination can be expected due to the low cross-polarization level for each radiation pattern.

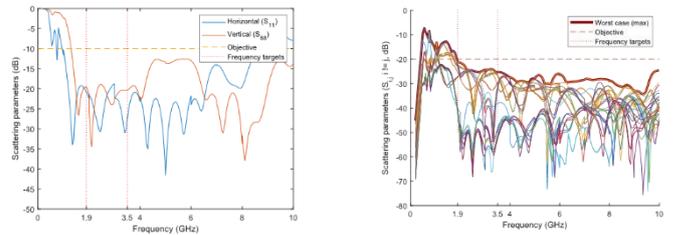

Figure 9. (a) Reflection coefficient $S_{11}$ (horizontal element) and $S_{88}$ (vertical element); (b) Isolation coefficient between $S_{ij}$ ports

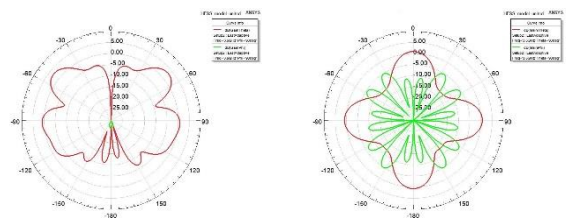

Figure 10. Gain(dB) in Co (red) and Cross (green) polarization for the equivalent electric dipole in E (left) and H plane (right)

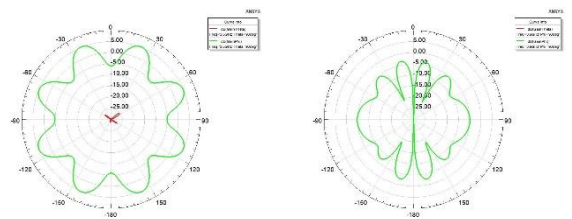

Figure 11. Gain(dB) in Co (green) and Cross (red) polarization for the equivalent magnetic dipole in the E (left) and H (right) plane

## B. Sub-6GHz Results

### 1) Uplink and Downlink Performance with VA and URA:
We evaluate VA and URA aided positioning performance with UL and DL reference signals under 0dB SNR situation. 20 reference symbols and 300MHz bandwidth with 15KHz spans are set for UL and DL two cases. The positioning simulation estimates AoA and Delay profile of UE by using SRS and PRS. Corresponding probability distribution functions (PDF) and Root Mean Square Error (RMSE) with EM and Subspace estimation methods are plotted in Fig. 12 (a). The mean value of EM method is slightly higher than subspace's. The subspace RMSE PDF is more concentrated at the mean value than EM. Curves of SRS and PRS are overlapped since they are using the same frequency domain and time domain resources (e.g., bandwidth, comb factors, and symbol numbers).

Fig. 12 (b) plots the mean RMSE with simulation repetition times for PRS and SRS cases. By comparing the curves of EM and subspace methods when using the same reference signal, EM curve convergences to 0.36 m and subspace convergences to 0.3 m, which is slightly lower than EM. Also, performance of PRS and SRS are the same.

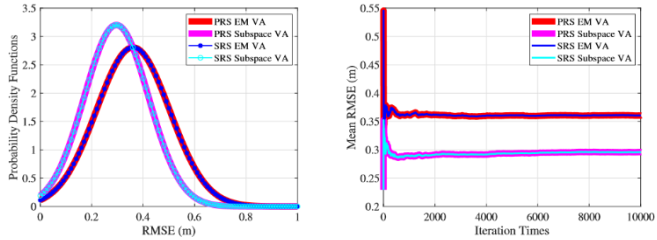

Figure 12. (a) RMSE PDF Plots; (b) Mean RMSE VS Iteration Times

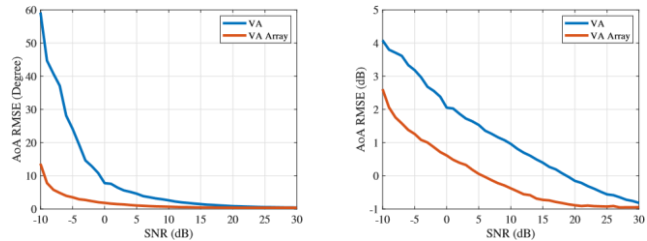

Figure 13. (a) AoA Estimation RMSE VS SNR, Linear Scale; (b) AoA Estimation RMSE VS SNR, Logarithm Scale.

### 2) Downlink Performance with VA Array:
Fig. 13 plots RMSE values of AoA estimation by using single VA and 2×2 VA array with half wavelength element spacing. Subspace method is used for searching AoA and the span is 1°. At the same RMSE value in Fig. 13 (a), VA array's curve requires around 13dB less SNR than VA's. In other words, VA array's curve has better estimation accuracy than VA with the same SNR value. This keeps true with higher SNR values (range from 15dB to 30dB) in Fig. 13 (b), logarithm scale plot. Intuitively, combination of antenna array structure with VA improve the AoA estimation accuracy even if the array size is as small as $2 \times 2$.

### 3) Imaging with 5G Signal:
We evaluate the 5G-aided imaging function by using SRS. The SRS signal takes up 150 MHz bandwidth with 15 kHz subcarrier spacing in the frequency domain, and it uses 2000 OFDM symbols in the time domain. Fig. 14 (a) and (b) plot reflection path of Muret airport and corresponding range Doppler processing output. By comparing two plots, the range-Doppler plot contains reflections marked by A to G and also mirrors of E and A. As SRS symbols are non-continuous among slots, the symbols which are not used by SRS will cause mirrors in the rangeDoppler plot result. Under this case, 91.84 m/s is the span between reflection points and their mirrors. To counter-effect such phenomenon, reducing the antenna beam or the flight speed are proper approaches. Based on simulation results, we can see reflections from major objects. This image will provide extra data source to improve the target detection and identification together with other sensor modals e.g. mmWave radar, optic sensors, or LiDAR. As a remark for futur work, it can be interesting to use this information (the main mutipath direction and location to use them in AOA and TOA algorithm.

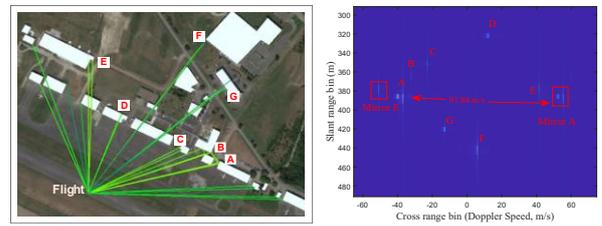

Figure 14. (a) Wireless InSite® Muret Ray Tracing Output. Flight transmitting SRS to gNB and objects illuminated by SRS signal are reflecting signal back to the flight. We set 24dBm TX power and 17.1dBi antenna gain. The received signal strength from every path is higher than the flight RX sensitivity −90dBm [23] [24]; (b) Range Doppler Plot. It contains seven reflection points (Marked as A to G) and two mirrors of A and E

## C. mmWave Results

Texas Instrument (TI)'s radar chips and evaluation boards - AWR1443, AWR1642, and AWR1843 - are used to assess mmWave technology as they provide small form factor boards allowing to retrieve directly on a PC raw ADC data from multiple Rx antenna (Fig. 15) (a). The assessment was done using basic unitary testing in a controller laboratory environment in addition to test scenarios performed at underground parking. A parking environment can be considered as a good baseline for validation with a lot of similarities with airport environment, mainly to detect vehicles and humans, and to validate the radar functional aspect (range, velocity, and angle information). A preliminary target detection is shown in Fig. 15 (b).

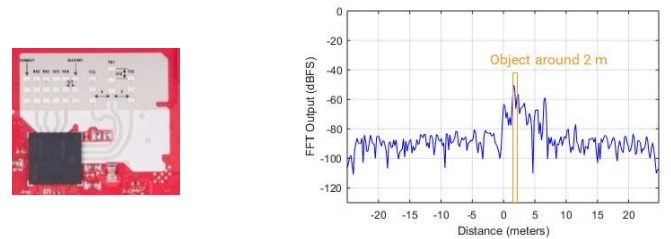

Figure 15. (a) AWR1843 board, TX and RX antennae details [25]; (b) 1D FFT amplitude profile (per chirp)

Additional testing in the next project phase at airport will allow to also test aircraft detection. The main challenge during the first project phase has been to control the software environments and to process the raw radar data. The first testing results confirmed the capability and the potential of the mmWave technology. The measured radar performance was aligned with the theoretical calculation, and the radar was mainly able to detect moving targets in a complex environment. In the next project phase, the main objective will be to consolidate a dataset for AI training using the different scenarios recording including the underground parking, and later on the Muret Airfield. This dataset will be used to train an AI model to identify target's type (person, vehicle, or aircraft).

## VI. CONCLUSION

In this paper, we described the motivation and collaborative approach for the elaboration of a new surveillance service for small and medium-sized airports, based on the 3D-VA enhanced 5G positioning and fusion of sub-6GHz 5G signal and mmWave sensing. The detailed design procedure of 3D-VA was described and optimized for the sub-6GHz 5G spectrum with preliminary measurements of S-parameter and radiation pattern. The angle-and-delay estimation algorithms, as well as the related positioning strategy were shown. We also showed the enhancement effect of the 3D-VA for the angle estimation. The working principle of the FMCW mmWave and the experimental results based on the TI AWR1843 were also shown. Based on the preliminary results of 3D-VA, angledelay estimation algorithms, sub-6GHz and mmWave sense, we showed that the selected techniques are competent for the roles in the original hypothesis. To further prove the benefits of the collaborative approach, the following work will be implemented: i), experimental 3D-VA angle estimation performance evaluation; ii), verification of the angle-delay algorithms and SAR capability with practical 5G radio samples; iii), intelligent target recognition with Doppler signature of mmWave radar; and iv), joint target registration based on sub-6GHz SAR image and mmWave radar detection.

### ACKNOWLEDGMENT

This work was supported by SESAR Joint Undertaking (SJU) in project NewSense (grant 893917). The opinions expressed herein reflect the authors' view only. Under no circumstances shall the SJU be responsible for any use that may be made of the information contained herein. This work was also partly supported by the Academy of Finland, under the projects ULTRA (328226) and ACCESS (339519).

### REFERENCE

[1] G. Fodor, J. Vinogradova, P. Hammarberg, K. K. Nagalapur, Z. T. Qi, H. Do, R. Blasco, and M. U. Baig, "5g new radio for automotive, rail, and air transport," IEEE Communications Magazine, vol. 59, no. 7, pp. 22–28, 2021.

[2] H. Wymeersch, G. Seco-Granados, G. Destino, D. Dardari, and F. Tufvesson, "5g mmwave positioning for vehicular networks," IEEE Wireless Communications, vol. 24, no. 6, pp. 80–86, 2017.

[3] B. Sun, B. Tan, W. Wang, and E. S. Lohan, "A comparative study of 3d ue positioning in 5g new radio with a single station," Sensors, vol. 21, no. 4, 2021.

[4] O. Kanhere, S. Goyal, M. Beluri, and T. S. Rappaport, "Target localization using bistatic and multistatic radar with 5g nr waveform," in 2021 IEEE 93rd Vehicular Technology Conference (VTC2021-Spring), pp. 1–7, 2021.

[5] W. Kim, H. Cho, J. Kim, B. Kim, and S. Lee, "Yolo-based simultaneous target detection and classification in automotive fmcw radar systems," Sensors, vol. 20, no. 10, 2020.

[6] J. Jakobi, M. Roeder, M. Biella, and J. Teutsch, "Economic aspects of advanced surface movement guidance and control systems," 09 2009.

[7] M. C. Vanderveen, A. . Van der Veen, and A. Paulraj, "Estimation of multipath parameters in wireless communications," IEEE Transactions on Signal Processing, vol. 46, no. 3, pp. 682–690, 1998.

[8] J. Fessler and A. Hero, "Space-alternating generalized expectationmaximization algorithm," IEEE Transactions on Signal Processing, vol. 42, no. 10, pp. 2664–2677, 1994.

[9] EUROCAE, ED-117A – MOPS for Mode S Multilateration systems for use in A-SMGCS, 2016.

[10] EUROCAE, ED-116 – MOPS for Surface Movement Radar Sensor Systems for use in A-SMGCS, Jan. 2004.

[11] F. B. Slama and et al., D1.1-Operational, security and safety preliminary requirements analysis. NewSense, Oct. 2021.

[12] 3GPP, "3gpp tr 21.916 technical specification group services and system aspects release 16 description;," vol. 0, 09 2020.

[13] 3GPP, "3gpp tr 21.917 technical specification group services and system aspects; release 17 description," vol. 0, 11 2021.

[14] J. Duplouy, C. Morlaas, H. Aubert, P. Potier, and P. Pouliguen, "Radiation-pattern reconfigurable and wideband vector antenna for 3d direction finding," URSI Radio Science Bulletin, vol. 2020, no. 373, pp. 56–62, 2020.

[15] E. Dahlman, S. Parkvall, and J. Skold, "Chapter 24: Positioning," pp. 479–485, 2020.

[16] R. O. Schmidt, "Multiple emitter location and signal parameter estimation," Adaptive Antennas for Wireless Communications, vol. 34, no. 3, pp. 276–280, 1986.

[17] Y. Golovachev, A. Etinger, G. Pinhasi, and Y. Pinhasi, "The effect of weather conditions on millimeter wave propagation," international journal of circuits, systems and signal processing, pp. 690–695, 2019.

[18] X. Gao, G. Xing, S. Roy, and H. Liu, "Experiments with mmwave automotive radar test-bed," 2019 53rd Asilomar Conference on Signals, Systems, and Computers, Nov 2019.

[19] X. Gao, G. Xing, S. Roy, and H. Liu, "Ramp-cnn: A novel neural network for enhanced automotive radar object recognition," IEEE Sensors Journal, vol. 21, p. 5119–5132, Feb 2021.

[20] M. Ester, H. Kriegel, J. Sander, and X. Xu, "A density-based algorithm for discovering clusters in large spatial databases with noise," in KDD, 1996.

[21] D. Cha, S. Jeong, M. Yoo, J. Oh, and D. Han, "Multi-input deep learning based fmcw radar signal classification," Electronics, vol. 10, 2021.

[22] G. E. Smith, K. Woodbridge, and C. J. Baker, "Micro-doppler signature classification," in 2006 CIE International Conference on Radar, pp. 1–4, 2006.

[23] J. Tetazoo, "5g network rf planning – link budget basics." Available at https://www.techplayon.com/5g-network-rf-planning-linkbudget-basics (2021/06/15).

[24] CommScope, Base Station Antenna Data Sheet, 8 2021. Multibeam Antenna.

[25] T. Instruments, "Xwr1843 evaluation module (xwr1843boost) singlechip mmwave sensing solution user's guide, spruim4b," May 2020.

[26] G. Eason, B. Noble, and I. N. Sneddon, "On certain integrals of Lipschitz-Hankel type involving products of Bessel functions," Phil. Trans. Roy. Soc. London, vol. A247, pp. 529–551, April 1955. (references)

[27] J. Clerk Maxwell, A Treatise on Electricity and Magnetism, 3rd ed., vol. 2. Oxford: Clarendon, 1892, pp.68–73.

[28] I. S. Jacobs and C. P. Bean, "Fine particles, thin films and exchange anisotropy," in Magnetism, vol. III, G. T. Rado and H. Suhl, Eds. New York: Academic, 1963, pp. 271–350.

[29] K. Elissa, "Title of paper if known," unpublished.

[30] R. Nicole, "Title of paper with only first word capitalized," J. Name Stand. Abbrev., in press.

[31] Y. Yorozu, M. Hirano, K. Oka, and Y. Tagawa, "Electron spectroscopy studies on magneto-optical media and plastic substrate interface," IEEE Transl. J. Magn. Japan, vol. 2, pp. 740–741, August 1987 [Digests 9th Annual Conf. Magnetics Japan, p. 301, 1982].

[32] M. Young, The Technical Writer's Handbook. Mill Valley, CA: University Science, 1989.